\documentclass[twocolumn,showpacs,amsmath,pre]{revtex4}
\usepackage{graphicx}
\usepackage{amsmath}
\usepackage{amssymb}
\usepackage{natbib}
\usepackage{subfigure}
\usepackage{color} 

\begin{document}

\title{The statistics of frictional families}

\author{Tianqi Shen$^{1}$}
\author{Stefanos Papanikolaou$^{2,1}$}
\author{Corey S. O'Hern$^{2,1,3}$}
\author{Mark D. Shattuck$^{4,2}$}

\affiliation{$^{1}$Department of Physics, Yale University, New Haven,
Connecticut 06520-8120, USA}

\affiliation{$^{2}$Department of Mechanical Engineering \& Materials Science, Yale University, New Haven, Connecticut 06520-8260, USA}

\affiliation{$^{3}$Department of Applied Physics, Yale University, New Haven, Connecticut 06520-8120, USA}

\affiliation{$^{4}$Benjamin Levich Institute and Physics Department, The City College of the City University of New York, New York, New York 10031, USA}

\begin{abstract}
We develop a theoretical description for mechanically stable
frictional packings in terms of the difference between the total
number of contacts required for isostatic packings of frictionless
disks and the number of contacts in frictional packings,
$m=N_c^0-N_c$.  The saddle order $m$ represents the number of
unconstrained degrees of freedom that a static packing would possess
if friction were removed. Using a novel numerical method that allows
us to enumerate disk packings for each $m$, we show that the
probability to obtain a packing with saddle order $m$ at a given
static friction coefficient $\mu$, $P_m(\mu)$, can be expressed as a
power-series in $\mu$. Using this form for $P_m(\mu)$, we
quantitatively describe the dependence of the average contact number on
friction coefficient for static disk packings obtained from direct
simulations of the Cundall-Strack model for all $\mu$ and $N$.
\end{abstract}

\maketitle

Granular media are fascinating, complex materials that display gas-,
liquid-, and solid-like behavior depending on the boundary and driving
conditions.  Frictional forces are crucial for determining the
structural and mechanical properties of granular media in the
solid-like state~\cite{makse_compress}.  For example, friction plays an
important role in setting the angle of repose~\cite{angle}, determining
the width of shear bands in response to applied stress~\cite{shear,xu},
and enabling arches to form and jam hopper flows~\cite{pak}.

\begin{figure}
\includegraphics[scale=0.45]{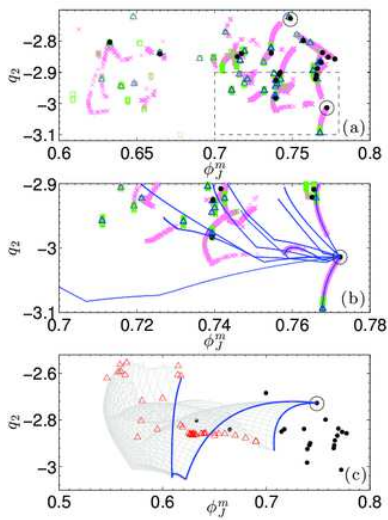}
\caption{(Color online) (a) The second invariant $q_2$ of the distance
matrix versus packing fraction at jamming onset $\phi_J^m$ for static
packings of bidisperse disks with $N=6$ generated using the
Cundall-Strack model for friction with $\mu=0$ (circles), $0.002$
(triangles), $0.02$ (squares), and $0.2$ (crosses).  Only packings
with saddle order $m=0$ and $1$ are shown. For $N=6$, there are
$N_s=20$ packings with $m=0$~\cite{gao}. (b) Close-up of boxed region
in (a) with $m=1$ packings (lines) generated using the `spring
network' method that originate from the circled $m=0$ packing. (c) All
$m=2$ packings ({\it i.e.} branch ${\cal A}$ shown as a gray mesh)
that are generated from the two highlighted families of first-order
saddle packings (dark lines) using the spring network method. $m=2$
packings with the same contact network as that for branch ${\cal A}$
generated using the Cundall-Strack method are also shown as
triangles.}
\label{invariant}
\end{figure}
   
For static packings of frictionless spherical particles, it is well
known that the minimum contact number required for mechanical
stability~\cite{gao} is $\langle z\rangle_{\rm min}^0 = 2N_c^0/N$,
where $N_c^0 = dN - d +1$ is number of contacts among $N$ particles in
the force-bearing backbone of the system and $d$ is the spatial
dimension.  However, at nonzero static friction coefficient $\mu$,
fewer contacts are required for mechanical stability with $N_c \ge
N(d+1)/2 -1 + 1/d$ and $\langle z \rangle^{\infty}_{\rm min} = d+1$ in
the large-$N$ and $\mu$ limits.

Several computational studies have measured the contact number as a
function of $\mu$ for packings of frictional disks and spheres using
`fast' compression algorithms that generate amorphous
configurations~\cite{silbert1,makse,makse1}.  In particular, these
studies find $\langle z \rangle= \langle z \rangle_{\rm min}^{0}=4$
and $\langle z \rangle_{\rm min}^{\infty}=3$ in the $\mu \rightarrow
0$ and $\infty$ limits, respectively, for bidisperse
disks~\cite{silbert,papanikolaou}.  For intermediate values of $\mu$,
$\langle z\rangle$ smoothly varies between $\langle z \rangle_{\rm
min}^{0}$ and $\langle z \rangle_{\rm min}^{\infty}$.  However, it is
not currently known what determines the contact number distribution
for each $\mu$ and form of $\langle z (\mu)\rangle$ for a given
packing preparation protocol.  The ability to predict the functional
form of the contact number with $\mu$ is important because $\langle z
\rangle$ controls the mechanical~\cite{somfai} and
vibrational~\cite{bertrand} properties of granular packings.

In this Letter, we develop a theoretical description for static
frictional packings at jamming onset in terms of their `saddle order,'
or the number of contacts that are missing relative to the isostatic
value in the zero-friction limit, $m=N_c^0-N_c$.  In contrast,
previous studies used $\mu \rightarrow \infty$ packings as the
reference~\cite{henkes}. Using a novel numerical procedure (the
`spring network' method) that allows us to enumerate packings for each
$m$ and molecular dynamics (MD) simulations of the Cundall-Strack
model~\cite{cundall} for frictional disks, we show that $m$
characterizes the dimension of configuration space that static
packings occupy.  Frictional packings with $m=1$ contacts form
one-dimensional lines in configuration space, packings with $m=2$
populate two-dimensional areas in configuration space, and packings
with larger $m$ form correspondingly higher-dimensional structures in
configuration space.  We assume that the probability for obtaining a
static packing with saddle order $m$ at a given $\mu$, $P_m(\mu)$, is
proportional to the volume occupied $V_m(\mu)$ by force- and
torque-balanced $m$th order saddle packings in configuration space.
We find that $P_m(\mu)$ can be written as a power-series in $\mu$,
$P_m(\mu) \sim a_m \mu^m/(1+ \sum_{i=1}^{N/2-1} a_i \mu^i)$, where
$a_m$ are the normalized coefficients of the power series.  Using this
form, we are able to quantitatively describe the dependence of the
average contact number on the friction coefficient for static disk
packings obtained from MD simulations of the Cundall-Strack model over
a wide range of $\mu$ and in the large system limit.

We generated static packings of bidisperse ($50$-$50$ mixtures of
particles with equal mass and diameter ratio $\sigma_1/\sigma_2=1.4$)
frictional disks in square cells with periodic boundary conditions
using two methods. First, we implemented a packing-generation
algorithm in which the system is isotropically compressed or
decompressed (followed by energy minimization) to jamming
onset~\cite{gao} at packing fraction $\phi_J^m$ that depends on the
saddle order. Pairs of overlapping disks $i$ and $j$ interact via
repulsive linear spring forces ${\vec F}^n_{ij}$ in the direction of
the center-to-center separation vector ${\vec r}_{ij}$.  We
implemented the Cundall-Strack model for the frictional
interactions. When disks $i$ and $j$ come into contact, a tangential
spring is initiated with a force ${\vec F}^t_{ij}$ that is
proportional to the tangential (perpendicular to ${\hat r}_{ij}$)
displacement $u^t_{ij}$ between the disks.  The tangential
displacement is truncated so that the Coulomb threshold, $|F^t_{ij}|
\le \mu |F^n_{ij}|$, is always satisfied.  When the disk pairs come
out of contact, we set $u^t_{ij}$ to zero.

\begin{figure}
\includegraphics[scale=0.55]{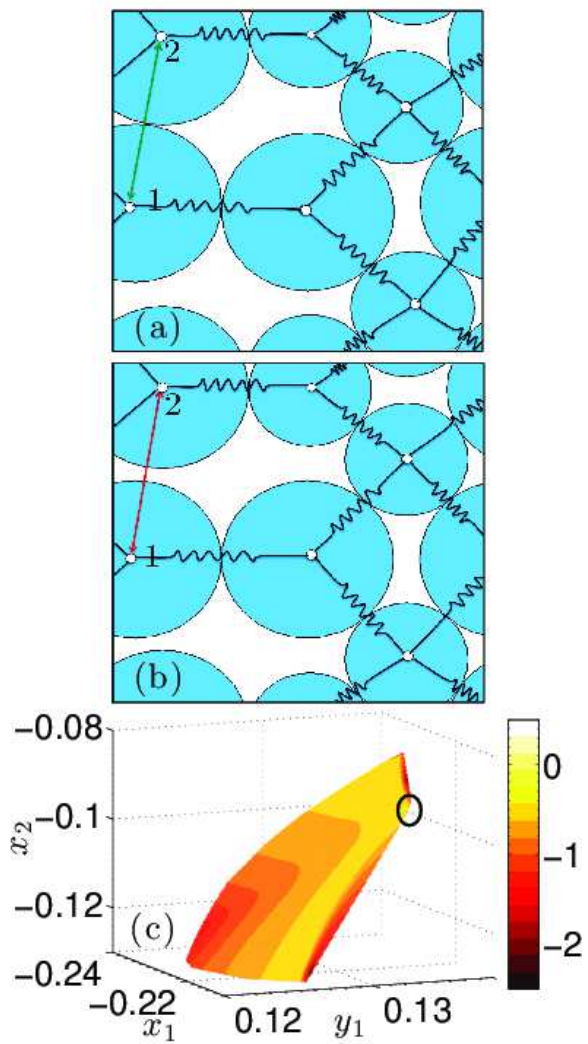}
\caption{(Color online) (a) Zeroth-order saddle ($m=0$) packing of
$N=6$ bidisperse frictionless disks with $N_c=N_c^{0}=11$
interparticle contacts.  To enumerate packings with $m=1$, the contact
between disks $1$ and $2$ is broken and constrained to have separation
(b) $r_{12}/\sigma_{12} = \lambda >1$, while the other $10$ contacts
are maintained (visualized as springs) at $r_{ij} = \sigma_{ij}$. The
successive compression and decompression packing-generation
process~\cite{gao} is performed with these
constraints to create an $m=1$ packing at $\phi_J^1$ with only $10$
contacts. This process is then repeated as a function of $\lambda$ and
for the other $10$ contacting particle pairs. (c) Contour plot of the
minimum friction coefficient $\mu_{\rm min}$ required to achieve force
and torque balance for a branch of $m=2$ packings in configuration
space spanned by the central position of the spring $(x_1, y_1)$
constraining the first broken contact and $x$-component of the spring
constraining the second broken contact $x_2$.  The $m=2$ branch of
packings emanates from the circled $m=0$ packing.  The color scale
for $\log \mu_{\rm min}$ increases from dark to light.}
\label{spring}
\end{figure}

We characterize each disk packing in configuration space by plotting
the second invariant $q_2 = (\operatorname{tr}^2(D) -
\operatorname{tr}(D^2))/2$ of the $N\times N$ distance matrix, $D_{ij}
= \sqrt{(x_i - x_j)^2 + (y_i-y_j)^2}$ versus $\phi_J^m$, where $x_i$
and $y_i$ are the $x$- and $y$-coordinates of particles $i$ and $j$.
(Note that $q_2$ is invariant to uniform translations and rotations,
as well as particle-label permutations, of the system.)  The plot of
$q_2$ versus $\phi_J^m$ in Fig.~\ref{invariant} (a) for packings with
$m=0$ and $1$ illustrates several important features. First, in the
$\mu \rightarrow 0$ limit, $m=0$ packings occur as distinct points in
configuration space (or $q_2$ versus $\phi_J^m$)~\cite{gao}. Second,
as $\mu$ increases, $m=1$ packings form one-dimensional lines in
configuration space that emanate from $m=0$ packings.  The $m=1$
packings that are stabilized at low $\mu$ are displaced in
configuration space from the $m=0$ packings as highlighted in
Fig.~\ref{invariant} (b).  In contrast, the packings that occur at
large $\mu$ approach the $m=0$ packings. Thus, we find that the
lengths of the $m=1$ lines increase with $\mu$. $m=2$
(Fig.~\ref{invariant} (c)) and higher-order saddle packings populate
areas and higher-order volumes in configuration space.

We also developed a numerical technique (`spring network' method) to
enumerate packings at each $m$. The method is best explained using an
example.  In Fig.~\ref{spring} (a), we show an $m=0$ packing of $N=6$
frictionless disks with $N_c = N_c^0=11$ contacts, which corresponds
to the packing circled in the lower right corner of the $q_2-\phi_J^m$
plane in Fig.~\ref{invariant} (a) and (b).  To systematically generate
$m=1$ packings with $10$ contacts, we break one of the $11$ contacts
in this packing ({\it e.g.} the contact between disks $1$ and $2$ in
Fig.~\ref{spring} (a)) and constrain its separation to be
$r_{12}/\sigma_{12} = \lambda > 1$, while the other contacts are
constrained to be $r_{ij}=\sigma_{ij}$. With these constraints and as
a function of $\lambda$, we implement the successive compression and
decompression packing-generation algorithm~\cite{gao} to find packings
at jamming onset, $\phi_{J}^1$.  This procedure is repeated for each
of the $10$ other contacts in the packing in Fig.~\ref{spring} (a) to
yield the $N_c^0=11$, $m=1$ branches in Fig.~\ref{invariant} (b), and
then for each of the $m=0$ packings. As shown in Fig.~\ref{invariant} (b), we
find overlap between the $m=1$ branches from the 
spring network method and the
$m=1$ packings generated from simulations of the Cundall-Strack
model. $m=2$ and higher-order saddle packings can be obtained using a
similar procedure, except multiple contacts are broken, as shown in
Fig.~\ref{invariant} (c). Thus, a family emanates from each $m=0$
packing with $N_b(N,m)$ branches with dimension $m$ in configuration
space for each saddle order.

\begin{figure}
\includegraphics[scale=0.37]{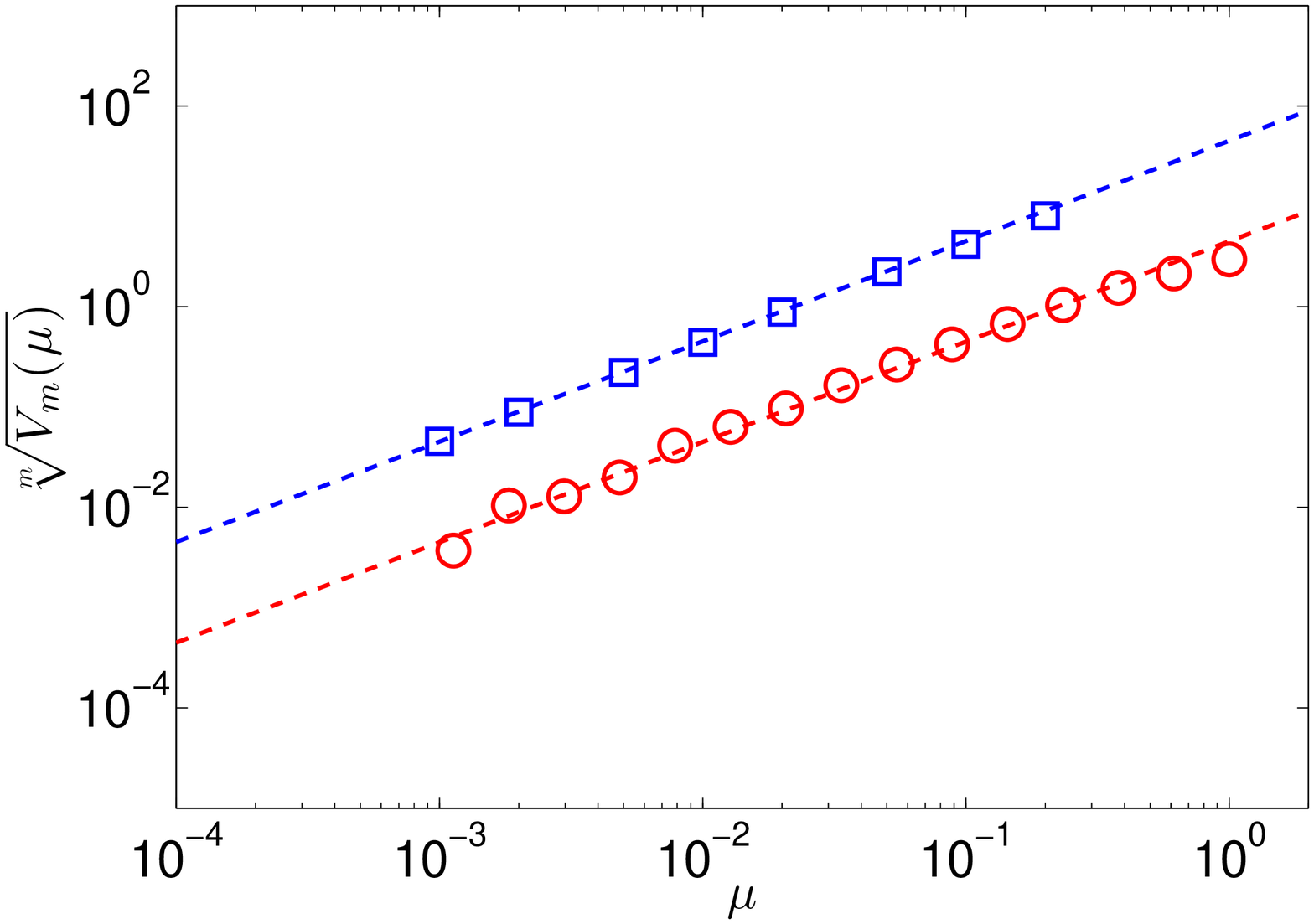}
\caption{(Color online) $m$th root of the volume $V_m(\mu)$ in configuration
space of the collection of $m$th order saddle packings ($m=1$,
squares; $m=2$, circles) that are stabilized by a friction coefficient $\le
\mu$. The dashed lines have slope $1$.}
\label{linear}
\end{figure}

We are interested in determining the probability $P_m(\mu)$ to obtain
an $m$th order saddle packing at a given $\mu$ averaged over families.
We will assume that $P_m(\mu)$ is proportional to the volume
$V_m(\mu)$ in configuration space of the $m$th order saddles that can
be force- and torque-balanced with friction coefficient $\mu$ or less:
\begin{equation}
P_m(\mu) \propto V_m(\mu) \delta^{2N - 1 - m}, 
\label{volume}
\end{equation} 
where $\delta$ is a lengthscale required to make Eq.~\ref{volume}
dimensionally correct.  To measure $V_m(\mu)$, we first employ the
spring network method to generate a grid of points for each branch of
saddles of order $m$.  At each grid point characterized by $(x_1, y_1,
\dots,x_m, y_m)$, we determine the minimum friction coefficient
$\mu_{\rm min}(x_1,y_1,\ldots,x_m, y_m)$ required to achieve
mechanical equilibrium for that configuration, using Monte-Carlo moves
to search the null-space of the force- and torque-balance
matrix~\cite{shaebani}. The allowed configuration volume is determined
by integrating over the $\mu_{\rm min}$ contour (as shown in
Fig.~\ref{spring} (c)) such that $\mu \le \mu_{\rm
min}(x_1,y_1,\ldots,x_m, y_m)$ for a given $m$th order branch.  The
total admissible volume in configuration space $V_m(\mu)$ is obtained
by summing the volumes over all $m$th order branches.  We show in
Fig.~\ref{linear} that $V_m^{1/m}(\mu)$ scales linearly with $\mu$ for
$m=1$ and $2$.

\begin{figure}
\includegraphics[scale=0.44]{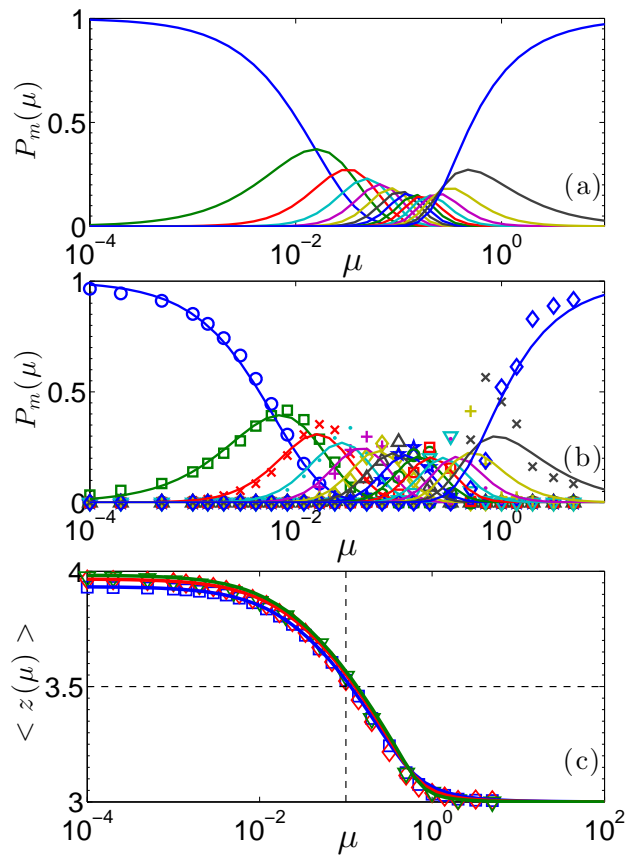}
\caption{(Color online) (a) The probability $P_m(\mu)$ to obtain a
packing with $m=0$, $1$, $\ldots$, $15$ (peak moving from left to
right) as a function of friction coefficient $\mu$ predicted by
Eq.~\ref{probability_eq} with $c_m(N) = 1$ and $N=30$.  (b) $P_m(\mu)$
for $m=0$ (circles), $1$ (squares), $2$ (exes), $\ldots$, $15$
(diamonds) moving from left to right from the Cundall-Strack model
(symbols) for $N=30$ and fits to Eq.~\ref{probability_eq} (lines) with
$c_m(N)$ given by Eq.~\ref{guess}. (c) $\langle z(\mu)\rangle$ for the
Cundall-Strack model for $N=30$ (squares), $64$ (diamonds), and $128$
(triangles) and accompanying fits to Eqs.~\ref{probability_eq} and~\ref{guess}
(lines).}
\label{probability}
\end{figure}

Thus, from the results in Fig.~\ref{linear}, we postulate the
following form for the normalized probability for an $m$th order
saddle:
\begin{equation}
P_m(\mu) = \frac{A_m \mu^m}{\sum_{m=0}^{m_{\rm max}} A_m \mu^m} = 
\frac{a_m \mu^m}{1+\sum_{m=1}^{m_{\rm max}} a_m \mu^m},
\label{probability_eq}
\end{equation}
where $a_m=A_m/A_0$ and the highest order saddle is $m_{\rm max} = N/2
- 1$ in 2D.  The coefficients $A_m = c_m(N) N_s(N) N_b(N,m)
\delta^{2N-1-m}$, where $N_s(N)$ is the number of $m=0$ packings for a
given $N$ and the normalized coefficients $a_m = c_m(N) N_b(N,m)
\delta^{-m}$.  A reasonable estimate for the number of branches stemming
from each $m=0$ packing is the $m$ permutations of $N_c^0$, $N_b(N,m)
= C^{N_c^0}_m$.  We show in Fig.~\ref{probability} (a) that
Eq.~\ref{probability_eq} with $c_m(N)=1$ and $\delta = 1$ yields
qualitatively correct results for the measured probabilities to obtain
a given $m$th order saddle from MD simulations of the Cundall-Strack
model for $N=30$ (Fig.~\ref{probability} (b)).  For example, zeroth
order saddles are most highly probable for small $\mu < 10^{-2}$, and
the highest order saddles are most probable for $\mu > 1$.  However,
as shown in Fig.~\ref{probability} (b), we obtain a much better fit to
the data from the Cundall-Strack model using
\begin{equation}
c_m(N) = \exp[-m[m-(m_{\rm
      max}+1)]/(m_{\rm max}+1)],
\label{guess} 
\end{equation}
which indicates an excess of $m$th order
saddles for small $m$ that likely originates from rattler particles
that join the force-bearing network during compression.

The average contact number can be obtained by calculating $\langle
z\rangle = (N_c^{0} - \langle m\rangle)/N$.  Our strategy is to use
the results for small systems ({\it i.e.} $N=30$) to predict
$P_m(\mu)$ and $\langle z\rangle$ versus $\mu$ for large $N$.  We show
in Fig.~\ref{probability} (c) that the predictions from
Eqs.~\ref{probability_eq} and~\ref{guess} agree quantitatively with
$\langle z(\mu)\rangle$ with no additional parameters for $N=64$ and
$128$ obtained from the Cundall-Strack model. Thus, we have developed
a method to calculate $\langle z(\mu) \rangle$ for large systems by
enumerating frictional families in small systems.

The calculations presented in this Letter provide a framework for
addressing several important open questions related to frictional
packings.  For example, why does the crossover from the low- to
high-friction limits in the average contact number and packing
fraction occur near $\mu* \approx 10^{-2}$ for
disks~\cite{papanikolaou,silbert} compared to $10^{-1}$ for
spheres~\cite{silbert,makse} for fast compression algorithms. In
addition, using the methods described above, we will be able to
determine how the crossover from low- to high-friction behavior
depends on the compression rate and degree of thermalization. Such 
calculations are crucial for developing the ability to design granular 
assemblies with prescribed structural and mechanical properties. 

We acknowledge support from NSF Grant No. CBET-0968013 (MS) and DTRA
Grant No. 1-10-1-0021 (CO and SP).  This work also benefited from the
facilities and staff of the Yale University Faculty of Arts and
Sciences High Performance Computing Center and NSF Grant
No. CNS-0821132 that partially funded acquisition of the computational
facilities.

\end{document}